\documentclass[onecolumn,amsmath,amssymb]{qm2008_abs}
\usepackage{graphicx}% Include figure files
\usepackage{dcolumn}% Align table columns on decimal point
\usepackage{bm}% bold math
\usepackage{units}
\newcommand{\bfig}[1][ht!]{\begin{figure}[#1] \begin{center}}
\newcommand{\efig}{\end{center} \end{figure}}
\newcommand{\bq}{\begin{equation}}
\newcommand{\eq}{\end{equation}}
\newcommand{\bqq}{\begin{eqnarray}}
\newcommand{\eqq}{\end{eqnarray}}

\begin{document}

\title{{\Large Measuring the Charged Particle Multiplicity with ALICE}}% Force line breaks with \\

\bigskip
\bigskip
\author{\large J. F. Grosse-Oetringhaus\footnote{Electronic address: \texttt{Jan.Fiete.Grosse-Oetringhaus@cern.ch}} for the ALICE collaboration}

\affiliation{CERN, 1211 Geneva 23, Switzerland and Institut f\"ur Kernphysik, M\"unster, Germany}
\bigskip
\bigskip

\begin{abstract}
\leftskip1.0cm
\rightskip1.0cm

The charged particle multiplicity distribution is one of the first measurements that ALICE will be able to perform. The knowledge of this basic property at a new energy is needed to configure Monte Carlo generators correctly with the aim of understanding the background of other, especially rare, processes including new physics. It allows to study the scaling behaviour and to verify model predictions. The unfolding of the measurement is a non-trivial task due to the finite precision and acceptance of the detector. Solutions are based on $\chi^2$ minimization or iteratively using Bayes' theorem. Both approaches to unfold the spectrum are presented. Furthermore, the capabilities of the SPD fast OR trigger are shown that enable physics at very high multiplicities.

\end{abstract}

\maketitle

\section{Introduction}

	A Large Ion Collider Experiment (ALICE) \cite{alice} will take data in p+p ($\sqrt{s} = \unit[14]{TeV}$) and Pb+Pb ($\sqrt{s_{\textnormal{NN}}} = \unit[5.5]{TeV}$) collisions. The charged particle multiplicity distribution $P(N)$, i.e. the probability $P$ for a collision with the charged particle multiplicity $N$, is one of the first measurements that can be performed.
	The knowledge of this basic property at a new energy is needed to tune Monte Carlo generators with the aim of understanding the background of other, especially rare, processes including new physics. It allows to study its scaling behavior and to verify model predictions.

	The multiplicity distribution of p+p($\bar{\mbox{p}}$) collisions at low energy is well-described by KNO scaling \cite{kno_scaling}, that is broken at SPS (Super Proton Synchrotron) energies ($\sqrt{s} = \unit[540]{GeV}$) \cite{kno_sps}.
	Other models and descriptions use negative binomial distribution (NBD) \cite{binom}, a two component approach \cite{twocomponent} or consider multiple parton interactions \cite{e735_nbd}.
	Only few predictions for LHC's energy realm exist, e.g. \cite{kaidalov_qcd200} (based on the Quark-Gluon String Model) and \cite{twocomponent_prediction} (based on the combination of two NBDs).

\begin{figure}[t!]
		\begin{minipage}[t]{0.46\linewidth}
			\centering
			\includegraphics[width=\linewidth,trim=0 20 0 20,clip=true]{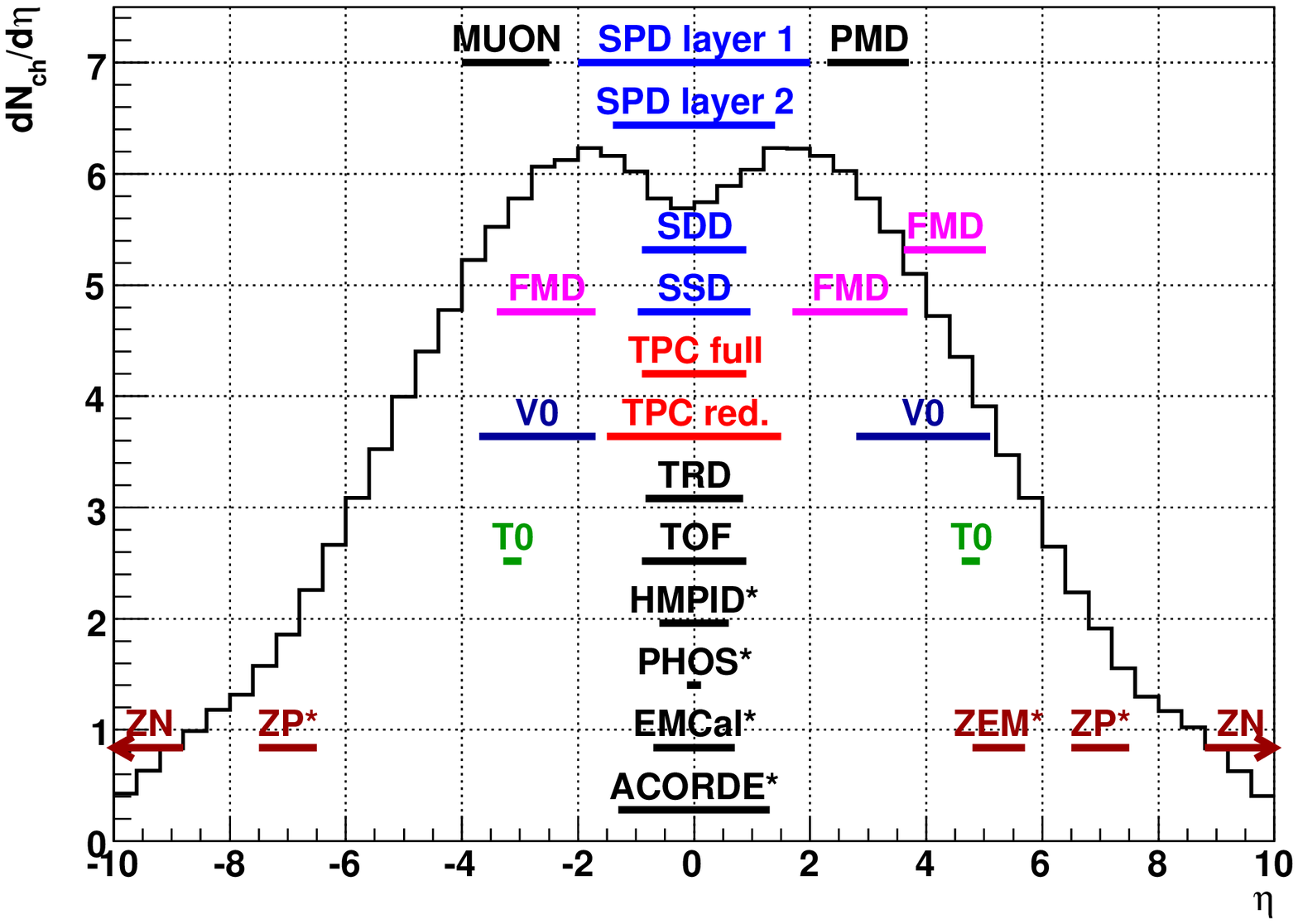}
			\caption{The Figure shows the acceptance in $\eta$ of all ALICE subdetectors overlayed with a $dN_{ch}/d\eta$ prediction by Pythia \cite{pythia} for $\sqrt{s} = \unit[14]{TeV}$. All subdetectors have full coverage in azimuth except the ones marked with an asterisk.}
			\label{fig_phasespace}
		\end{minipage}
		\hspace{0.4cm}
		\begin{minipage}[t]{0.46\linewidth}
			\centering
			\includegraphics[width=\linewidth,trim=0 20 0 20,clip=true]{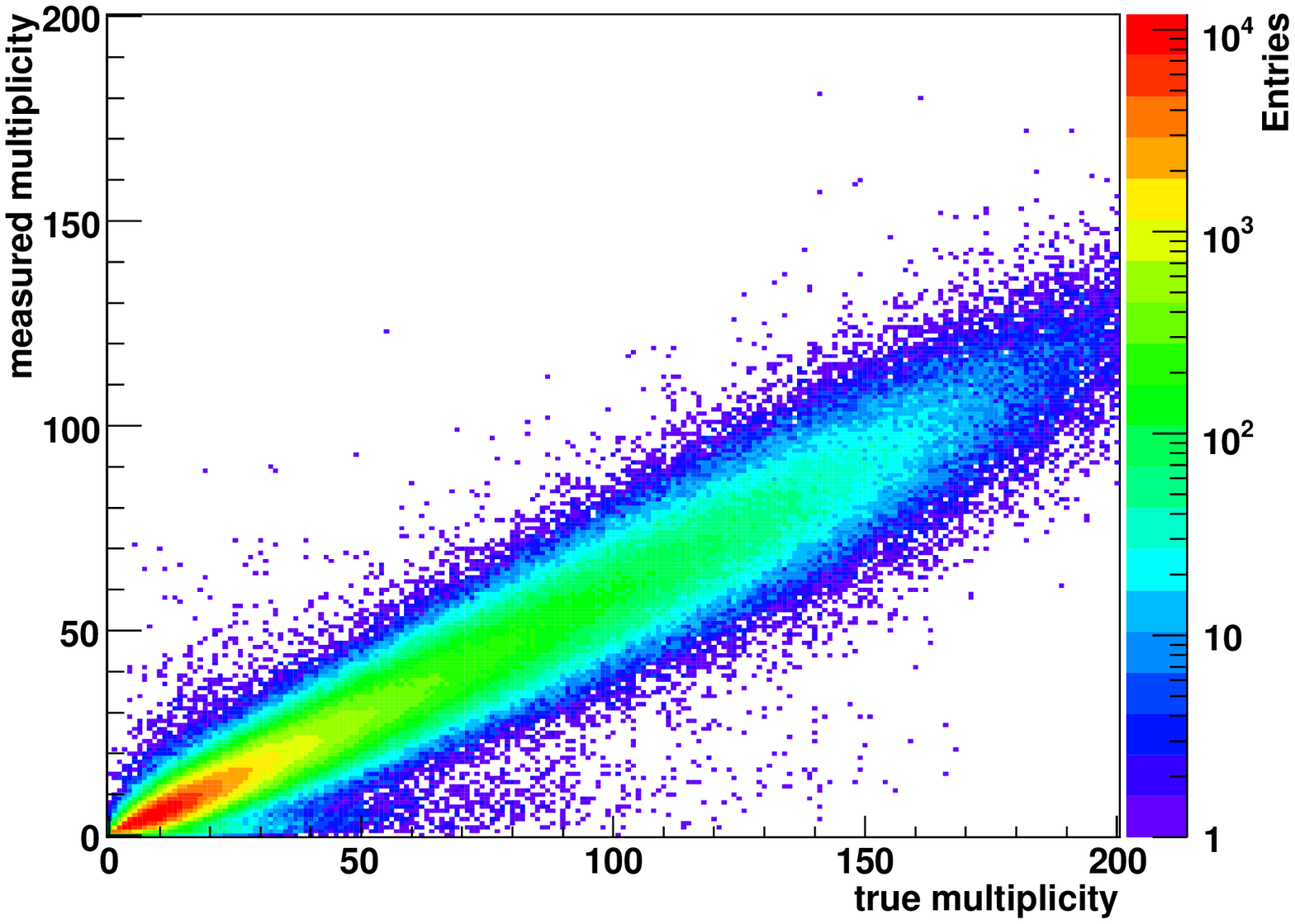}
			\caption{The Figure shows the detector response of the SPD, i.e. reconstructed tracklets vs. the number of generated primary particles in $|\eta| < 2.0$. Only events with a $z$ vertex position in $\pm \unit[10]{cm}$ are considered.}
			\label{responsematrix}
		\end{minipage}
		\vspace{-0.2cm}
	\end{figure}

	ALICE can measure the multiplicity distribution in a broad region of phase space. The region of $-3.4 < \eta < 5.1$ can be accessed combining the measurement of the Forward Multiplicity Detector (FMD) and the Silicon Pixel Detector (SPD). Figure \ref{fig_phasespace} shows the coverage in $\eta$ of the ALICE subdetectors.
	The analysis presented in these proceedings focuses on the measurement in the central region, using the information from the SPD in $|\eta| < 1.4$ and the measurement with the Time Projection Chamber (TPC) ($|\eta| < 0.9$). The latter can be used as a cross-check.

\section{Analysis}

	From all measured events only events that have a reconstructed primary vertex are considered for the analysis. This allows to select events from a given vertex range. Furthermore, the vertex is used to form tracklets in the SPD and to select good-quality TPC tracks.

	For each event the tracks that have a reasonable quality are counted. This step results in a measured multiplicity spectrum. Subsequently, this spectrum has to be unfolded in order to obtain the best estimate of the true multiplicity distribution (called \textit{true} spectrum in the following). This is non-trivial due to the fact that events with different true multiplicities contribute to the same measured multiplicity. The unfolding leads to the true spectrum -- the multiplicity distribution of charged primary particles -- for the events that have been triggered and have got a reconstructed vertex. Finally, this spectrum needs to be corrected for the bias introduced by the vertex reconstruction requirement and the trigger.

	\section{Unfolding}

	The response of the detector can be described by a matrix $R_{mt}$. It gives the conditional probability that
	a collision with a true multiplicity $t$ is measured as an event with the
	multiplicity $m$. The response matrix is created from the full detector simulation \cite{aliroot} (see Figure~\ref{responsematrix} for the response of the SPD).

	Given a true spectrum $T$, the measured spectrum $M$ can be calculated by $M = R T$.
	Therefore, given a measured spectrum, the true spectrum is formally calculated as follows $T = R^{-1} M$. However, $R^{-1}$ cannot be calculated in all cases, because $R$ may be singular. Even if $R$ can be inverted, the result obtained contains strong oscillations (due to statistical fluctuation caused by the limited number of events used to create the response matrix) \cite{blobel_unfolding}. Alternatives to simple inversion are Bayesian unfolding and $\chi^2$ minimization, introduced in the following paragraphs.

	\medskip

	\textbf{Bayesian Unfolding} is an iterative procedure based on Bayes' theorem which describes the definite relationship between the following two conditional probabilities: the probability of an event A ($m$ measured particles) conditional on another event B (true multiplicity $t$) and the probability of B conditional on A. Applying the theorem on the problem at hand results in \cite{blobel_unfolding, bayesian_unfolding}:

	\begin{minipage}{0.95\linewidth}
		\begin{minipage}[t]{0.48\linewidth}
			\centering
			\bq
				\tilde{R}_{tm} = \frac{ R_{mt} \cdot P_t }{ \sum\limits_{t'} R_{mt'} P_{t'} } \label{bayesian1}
			\eq
		\end{minipage}
		\begin{minipage}[t]{0.48\linewidth}
			\centering
			\bq
				U_t = \frac{1}{\epsilon_t} \sum\limits_{m} M_m \tilde{R}_{tm} \label{bayesian2}
			\eq
		\end{minipage}
		\vspace{0.2cm}
	\end{minipage}

\begin{figure}[t!]
		\begin{minipage}[t]{0.46\linewidth}
			\centering
			\includegraphics[width=\linewidth,trim=10 30 30 20,clip=true]{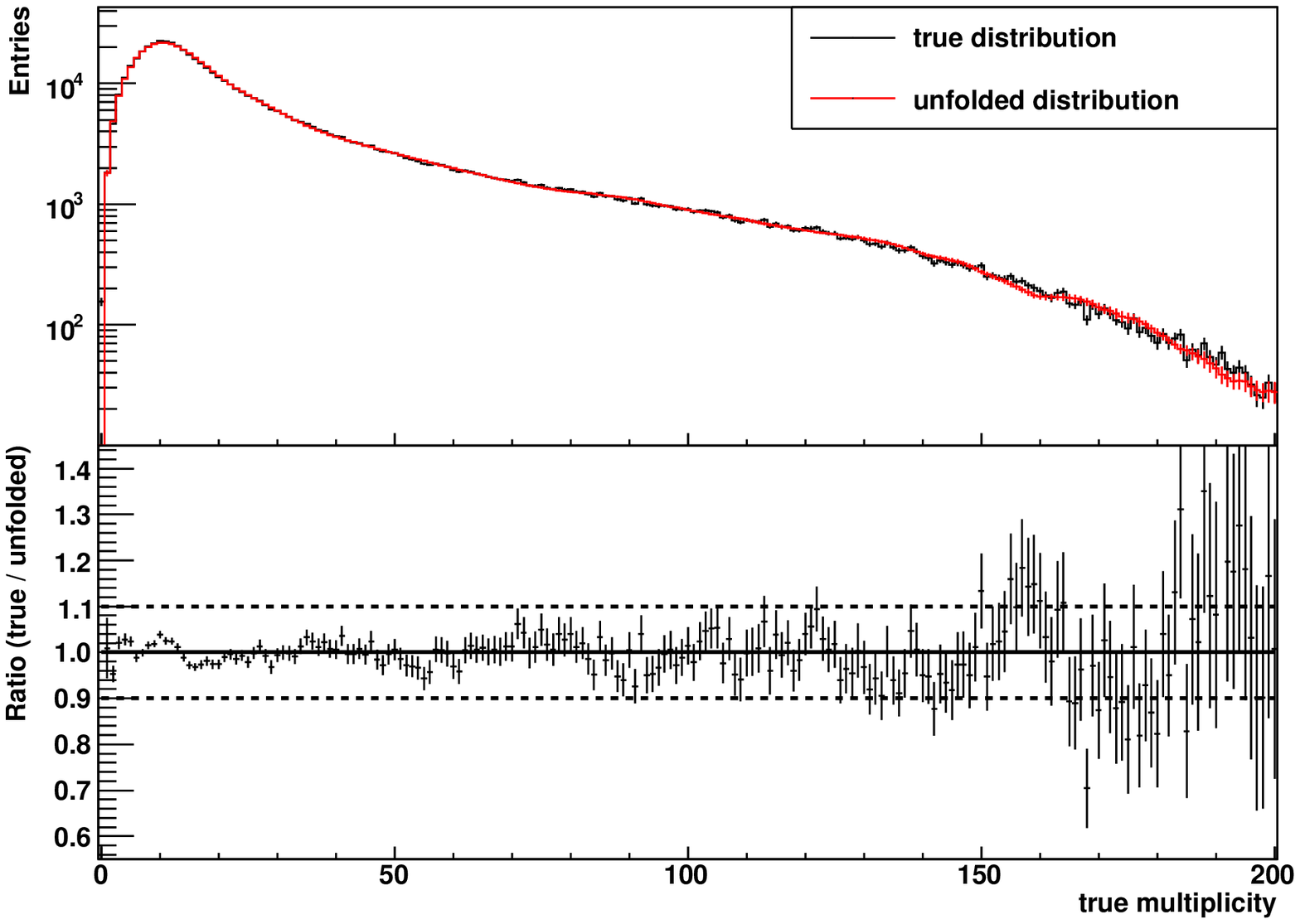}
		\end{minipage}
		\hspace{0.4cm}
		\begin{minipage}[t]{0.46\linewidth}
			\centering
			\includegraphics[width=\linewidth,trim=10 30 30 20,clip=true]{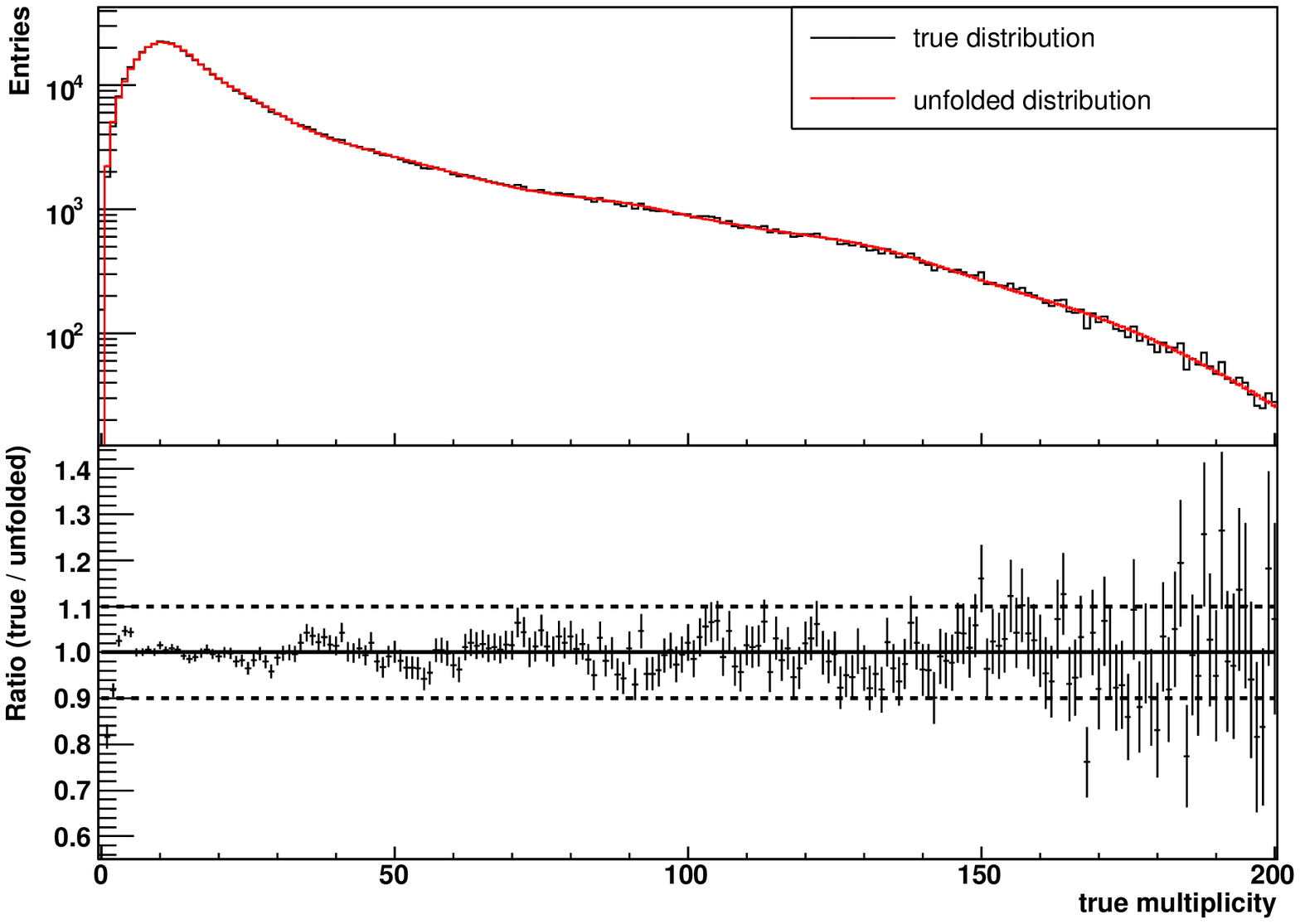}
		\end{minipage}
		\caption{Multiplicity distribution obtained with Bayesian unfolding (left panel) and $\chi^2$ minimization (right panel), overlayed with the true (input) distribution. The bottom part shows the ratio true/unfolded.}
		\label{fig_results}
	\end{figure}

	$R$ is the response matrix and $P$ the a-priori distribution of the true spectrum which can be chosen to be a flat distribution or like in the present analysis the measured spectrum. (\ref{bayesian1}) defines $\tilde{R}$, which is in literature also referred to as \textit{smearing matrix}. Subsequently, (\ref{bayesian2}) with $\tilde{R}$ and the measured spectrum $M$ allows to obtain $U$. $\epsilon_t$ is the efficiency to detect an event with a given true multiplicity. It is different from unity due to the vertex reconstruction and trigger efficiency. In the next iteration, $U$ is used as a-priori probability $P$. Optionally a smoothing is applied that reduces the influence of high-frequency fluctuations (e.g. a sliding average over three multiplicity bins). After about 10~--~100 iterations the result converges, $U$ approximates the true distribution. The left panel of Figure~\ref{fig_results} shows the result of a Bayesian unfolding after 100 iterations with smoothing.

	The strength of the smoothing as well as the number of iterations are free parameters that have been evaluated in detail in \cite{mult_note}.

	\medskip

	\textbf{$\chi^2$ Unfolding} is an approach to unfold the measured multiplicity distribution by the minimization of a $\chi^2$ function. Using the response matrix, this function gives a measure of how well a (guessed) unfolded spectrum describes the measured spectrum. A minimization program is used to find the unfolded spectrum that has the least $\chi^2$. With $e_m$ denoting the error on the measurement of $M$ and $U$ being the guessed spectrum, a suitable function is:
	\bq
		\chi^2(U) = \sum\limits_m \left(\frac{1}{e_m} \left(M_m - \sum\limits_t R_{mt} U_t\right)\right)^2 + \beta P(U). \label{chi2}
	\eq

	$P(U)$ is called regularization or penalty term. It only depends on the guess of the unfolded spectrum (and not on $R$ or $M$). It is a constraint that favors a certain shape of the unfolded spectrum and thus prevents the previously mentioned fluctuations. $\beta$ determines the weight that is given to the regularization with respect to the first term that governs the agreement with the measurement. A higher $\beta$ increases the value of the first term in (\ref{chi2}) because $U$ is more constrained. Its optimal value needs to be evaluated (see \cite{mult_note}); a reasonable value of $\beta$ adjusts the two terms in (\ref{chi2}) in a way that the introduced bias is negligible compared to the statistical error of the measurement.

	Many possibilities exist for the choice of the regularization $P$: these range from ``just'' requiring a smooth function to preferring a certain shape of the distribution. Generally, not too much knowledge about the distribution should be implied, otherwise the result is likely to look very similar to the required function and may mask details of the measurement. Various regularization functions have been evaluated in \cite{mult_note}, a suitable one is to require the least curvature:
	\bq
		P(U) = \sum\limits_t \left( \frac{U''_t}{U_t} \right)^2  = \sum\limits_t \left( \frac {U_{t-1} - 2U_t + U_{t+1}} {U_t} \right)^2 \label{linear}
	\eq

	An unfolding result is shown in the right panel of Figure~\ref{fig_results} that uses regularization (\ref{linear}) with $\beta = 10^4$.

\section{High Multiplicity Trigger}
\begin{figure}[t!]
		\begin{minipage}[t]{0.44\linewidth}
			\centering
			\includegraphics[width=\linewidth,trim=0 30 0 10,clip=true]{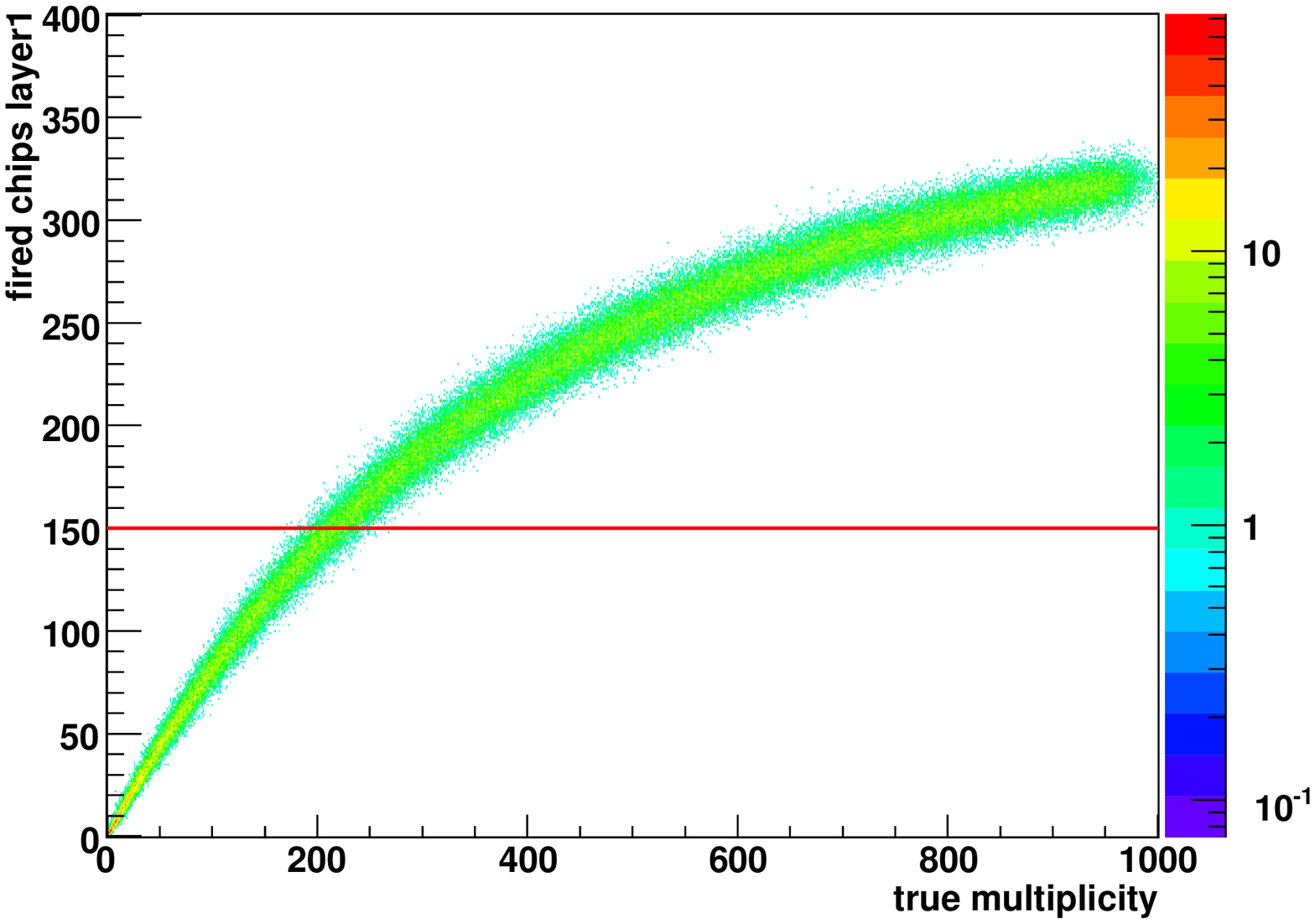}
		\end{minipage}
		\hspace{0.4cm}
		\begin{minipage}[t]{0.44\linewidth}
			\centering
			\includegraphics[width=\linewidth,trim=0 30 0 20,clip=true]{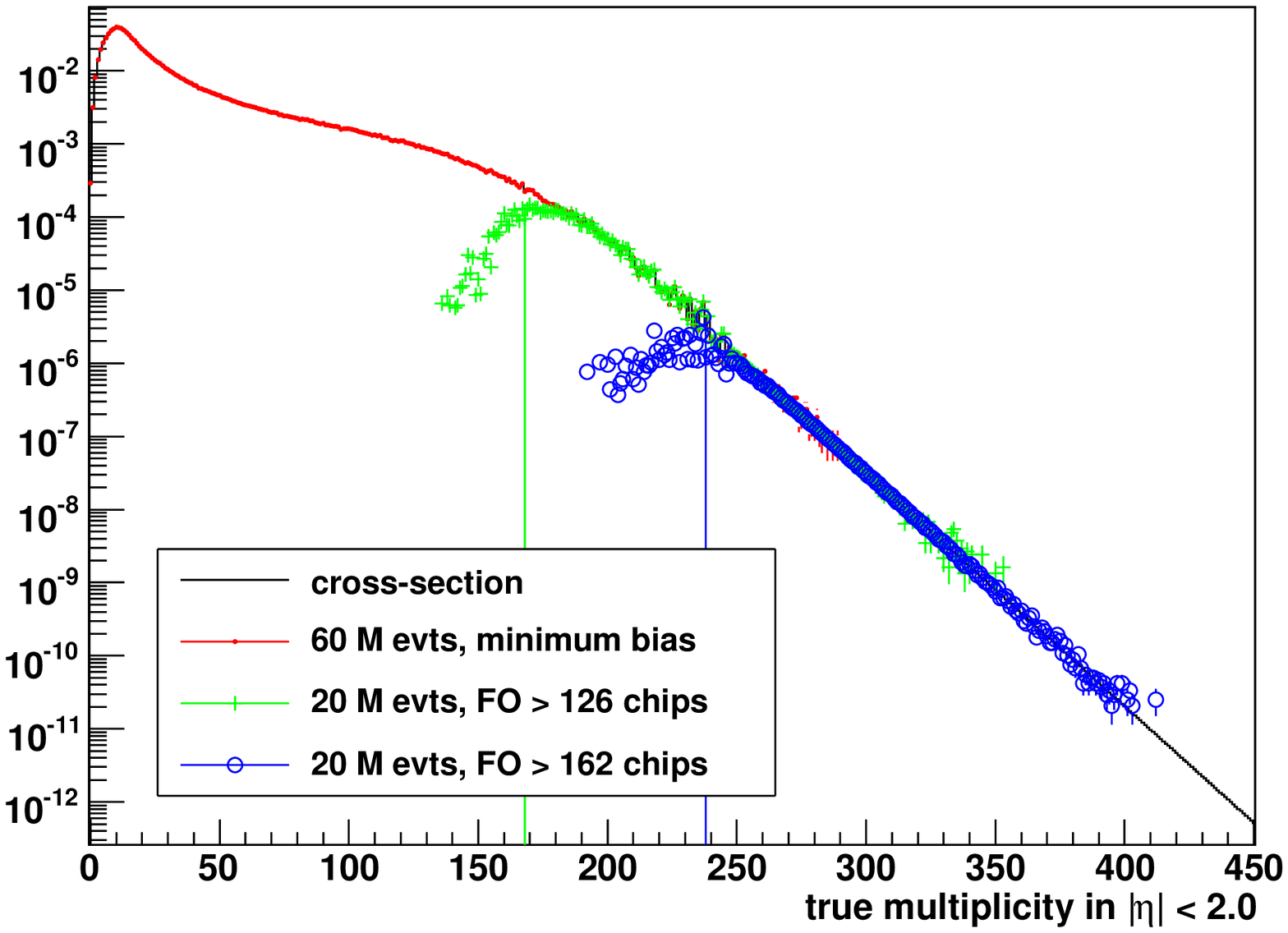}
		\end{minipage}
		\caption{Simulation results for the fast OR trigger. Left panel: response of the first layer; right panel: reach in multiplicity.}
		\label{fig_highmult}
	\end{figure}

	The multiplicity spectrum can be measured up to very high multiplicities by exploiting the SPD fast OR trigger. This trigger is based on the number of fired chips in the two layers of the SPD (radii of \unit[3.9]{cm} and \unit[7.6]{cm} from the beam line). A chip is fired if at least one particle deposits energy in one of the chip's attached pixels. In total 1200 chips produce a trigger at a rate of \unit[10]{MHz}. Trigger thresholds on the number of fired chips select high-multiplicity events; several thresholds can be configured simultaneously. The left panel of Figure~\ref{fig_highmult} shows for the first layer ($|\eta| < 2$) the number of fired chips vs. the event multiplicity together with a trigger treshold, set at 150 chips; the right panel shows a multiplicity distribution obtained by simulating a combination of the minimum bias trigger and two high multiplicity triggers.
	It reaches up to 400 charged particles, thus enabling to study e.g. the abundance of hyperons like the $\Xi$ at high multiplicities which provides insight in the hyperon scaling behavior with multiplicity.

\section{Summary}
	Two unfolding methods have been presented to obtain the multiplicity distribution. Either of them reproduces simulated spectra successfully. For the measurements, both methods should be used, as consistency check. Optimizations in both methods have been evaluated and the results described in these proceedings. More details and the study of the systematic effects that arise during this measurement can be found in \cite{mult_note}. A study performed on ALICE's high multiplicity trigger shows that this trigger promises interesting insight into the physics at high multiplicities.

\end{document}